\documentclass[aps, prl, twocolumn, superscriptaddress]{revtex4}

 \usepackage{amsmath, amssymb}
 \usepackage{bm, color}

\usepackage{graphicx}

\usepackage{natbib}

\newcommand{\be}{\begin{equation}}
\newcommand{\ee}{\end{equation}}

\newcommand{\vrr}{{\bf{r}}}

\newcommand{\vA}{{\bf{A}}}

\newcommand{\eps}{\epsilon}

\newcommand{\pa}{\partial}

\newcommand{\spp}{s_{++}}
\newcommand{\spm}{s_{+-}}

\begin{document}

\title{Phase soliton and  pairing symmetry of a two-band superconductor: \\ Role of the proximity effect}

\author{Victor Vakaryuk}
\email[]{vakaryuk@gmail.com}
\affiliation{Materials Science Division, Argonne National Laboratory, Argonne, Illinois 60439, USA}
\affiliation{Institute for Quantum Matter and Department of Physics \& Astronomy, The Johns Hopkins University, Baltimore, Maryland 21218, USA}

\author{Valentin Stanev}
\affiliation{Materials Science Division, Argonne National Laboratory, Argonne, Illinois 60439, USA}

\author{Wei-Cheng Lee}
\affiliation{Department of Physics, University of Illinois at Urbana-Champaign, Illinois 61801, USA}

\author{Alex Levchenko} 
\affiliation{Department of Physics and Astronomy, Michigan State University, East Lansing, Michigan 48824, USA}

\begin{abstract}


We suggest a mechanism which promotes the existence of a phase soliton -- topological defect formed in the relative phase of superconducting gaps of a two-band superconductor with $\spm$ type of pairing. This mechanism exploits the proximity effect with a conventional $s$-wave superconductor which favors the alignment of the phases of the two-band superconductor which, in the case of $\spm$ pairing, are $\pi$-shifted in the absence of proximity.  In the case of a strong proximity such effect can be used to reduce soliton's energy below the energy of a soliton-free state thus making the soliton thermodynamically stable. Based on this observation we consider an experimental setup,  applicable both for stable and metastable solitons, which can  be used to distinguish between $\spm$ and $\spp$ types of pairing in the iron-based multiband superconductors.

\end{abstract}



\maketitle

\emph{Introduction.} In the last decade multiband superconductivity became a central topic in condensed matter physics. The discovery of MgB$_2$  (Ref.~\onlinecite{MgB2}) and, more recently, of an entire family of iron-based high-temperature superconductors (Ref.~\onlinecite{Fe}) showed that it is not an exotic possibility, but a problem with enormous theoretical and practical relevance. Although the origin of superconductivity in MgB$_2$ and iron pnictides appears to be quite different (phonon-driven in the first, and purely electronic in the case of the latter), in all of these materials the presence of multiple gaps leads to important and far-reaching consequences. In fact, it has been argued that the multiband character of the Fermi surface is essential for the superconductivity in pnictides.    

It has been long known that two-band superconductors (SCs) have excitations associated with the fluctuations of the relative phase of the two gaps known as Leggett mode\cite{Leggett}. Only recently, however, has it been recognized that there are also topological excitations associated with the multiple gaps\cite{Tanaka:2001, Babaev:2002, Gurevich:2003, Babaev:2009, Lin:2011, Samokhin:2012}. In their simplest version these excitations can be thought of as  soliton-like domain walls between regions in which the relative phase differs by $2 \pi$.      

In this paper we argue that such solitons can be used to distinguish between the conventional $\spp$ and the more exotic $\spm$ pairing in which phases on the bands are $\pi$-shifted. Both states have been suggested as a possibility for iron pnictides. Despite being physically very different, distinguishing them experimentally is not a trivial task since they belong to the same symmetry class (for some interesting suggestions see e.g.~ Refs.~\onlinecite{Wu:2009} -- \onlinecite{Sudbo}). Here we elaborate a practical setup which  can be used to detect the difference between the two order parameters.  It utilizes the proximity effect which affects the existence of a soliton in a qualitatively different way depending on the relative sign between the gaps. 

\begin{figure}[b]
\vspace{-7pt}
\centering
	\includegraphics[scale=0.6]{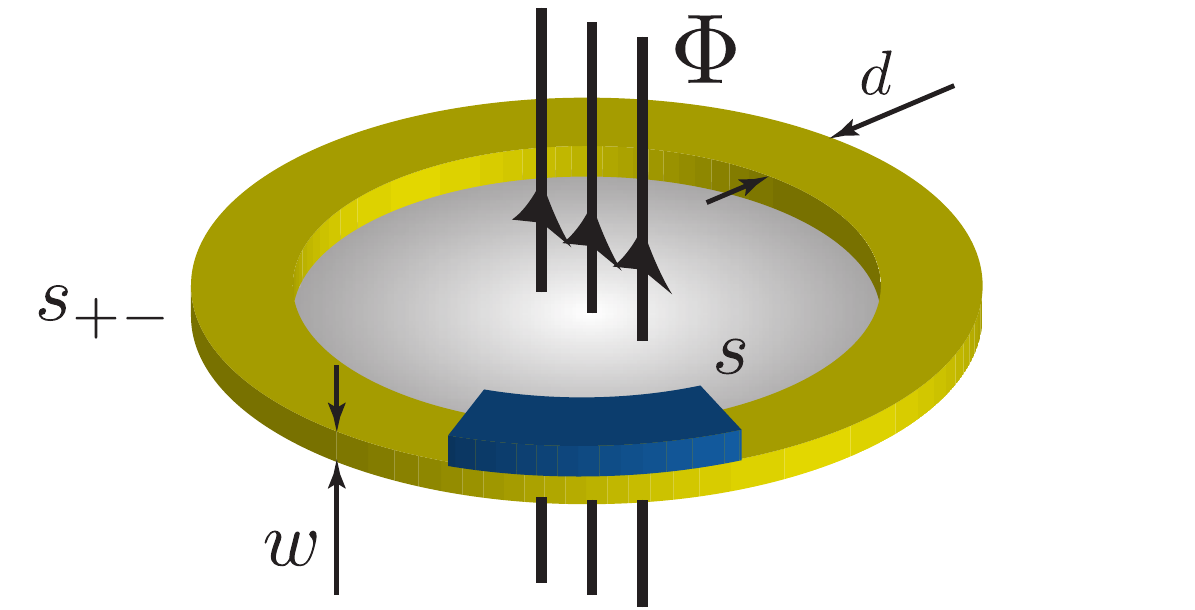}
	\caption{\label{patch} Proposed experimental setup which utilizes a proximity patch for the observation of a phase soliton.} 
	\label{patch}
\end{figure}

We consider a ring  made of a two-band SC, segment of which is covered with a stronger, higher $T_c$, $s$-wave superconductor (``proximity patch''). By the proximity effect the phases of the two bands will tend to align with the phase of the $s$-wave side (as discussed later), which reduces the energy of the soliton if  the two-band SC is in  $\spm$ state. For sufficiently strong proximity, the energy of the soliton can be brought below the energy of a vortex, thus making the former thermodynamically stable. We suggest to look for both equilibrium and metastable solitons by using magnetization measurements.

\emph{Description of the model.} Phenomenological description of a two-band superconductivity requires a two-component order parameter \cite{Tilley, Dao, Milo:Fluctuations}.
Let  $m_i$ be the effective mass of charge carriers in the $i$-th band and $\vA$ denote the vector potential.  Following early works  we consider a two-component Ginzburg-Landau (GL) model\cite{fn1} 
\vspace{-3pt}
\begin{eqnarray}
	F = \int_{\textsc{sc}} \!\!\!\! &\Big\{ & \!\!\! \sum_{i=1,2} \left[ \frac 1 {4m_i}  \left| \big( -i \hbar \bm \nabla - (2e/c) \vA \big) \psi_i\right|^2 
	+V_i(|\psi_i|) \right]
	\nonumber
	\\
	&-& \kappa (\psi_1 \psi_2^* + \psi_1^* \psi_2) \Big\} \, d \vrr
	\label{GL general}
\end{eqnarray} 
In this model each of the bands is assigned a complex gap $\psi_i =  \ |\psi_i| e^{i\theta_i}$ and is described by a standard GL functional with the potential energy term $V_i(|\psi_i|)$ (first line in Eq.~(\ref{GL general})).  Coupling between the bands is provided by a Josephson energy $E_{\rm J}$ which depends on their relative phase (second line in Eq.~(\ref{GL general})). In terms of the band phases this coupling can be written as $E_{\rm J} \equiv \int -2 \kappa |\psi_1| |\psi_2|  \cos(\theta_1 - \theta_2)$. Sign of the coupling constant $\kappa$ determines the relative phase of the bands in the ground state: $\kappa>0$ favors $\theta_1=\theta_2$ i.e.~the phases are aligned while $\kappa<0$ favors anti-aligned configuration $\theta_1 = \theta_2 +\pi$. These states  are often referred to as  $s_{++}$ and $\spm$ states, respectively.

%

Generally speaking the possibility of the relative phase dynamics depends on the strength of $\kappa$.  If $|\kappa|$ is very large then the phases are locked to each other and, as a result, act synchronously. However for weaker $|\kappa|$ behavior of the phases can be different. In particular one can envision a state in which phases of the order parameters on the two bands wind differently e.g.~host different number of vortices. A conceptually similar state -- half-quantum vortex -- has been recently observed in spin-triplet SC $\rm Sr_2 RuO_4$ where the role of the bands is played by the two spin components of charge carriers\cite{Jang:2011}.

To further elaborate on this idea we consider a ring of radius $R$, threaded by magnetic flux $\Phi$ (see Fig.~\ref{patch}). The height $w$ and the thickness $d$
 of the ring should be comparable or smaller than the characteristic lengthscale of the spatial variations of the order parameter\cite{Milo:2011}.  The single-valuedness condition requires both phases $\theta_i$ to wind only modulo $2\pi$ when taken around the ring.
For the relative phase $\alpha \equiv \theta_1 - \theta_2$ this implies that 
\vspace{-3pt}
\be
	\alpha (x + 2\pi R) = \alpha(x)+2\pi n_\alpha,
\label{winding}
\ee
where $x$ is the coordinate along the ring.
 For a regular vortex state $n_\alpha=0$ and the two phases have the same windings\cite{fn2}.
Importantly, there also exist states for which windings of $\theta_{1,2}$ around the ring are different corresponding to $n_\alpha= \pm 1, \pm 2 \ldots$ and non-zero winding in $\alpha$. Competition between the kinetic energy which tends to spread such winding and the Josephson coupling which, independent on its sign, tends to localize it, leads to the appearance of a spatially confined profile -- kink -- of the gradient of $\alpha$.  Such solution is called  phase soliton\cite{Tanaka:2001, Rajaraman}. 
Our aim is to show that the stability of the phase soliton can be affected by the proximity effect.

Proximity effects between conventional $s$-wave and  $\spm$ two-band SCs have been studied recently\cite{Ng, VS, AEK}.  While in the case of proximity between conventional (even multigap) SCs the effects are rather trivial, the proximity with $\spm$ order parameter provides for much richer physics. There is a number of anomalous features which can appear in such structures. In particular, it has been found that 
under some circumstances the phases on the two bands tend to align with the phase of the $s$-wave SC. As a results the relative phase between the bands deviates from the bulk $\pi$ value (note that the resultant state is 
complex and thus breaks time-reversal symmetry). If the proximity effect is sufficiently strong it is even possible to reduce the phase difference to zero and induce an $s_{++}$ state in the (originally) $\spm$ system.

It is intuitively clear that the proximity effect described above can affect soliton's stability. Indeed, in the case of $\spm$ symmetry the deviation of the relative phase from $\pi$ (configuration $\leftarrow \, \rightarrow$) --  essential requirement for the exis\-tence of the soliton -- is promoted in the region of the contact, thus lowering soliton's energy. It is also clear that if the contact is made between $\spp$ and $s$-wave materials this will, at best, preserve soliton's energy  since any deviation from $\uparrow \uparrow$ configuration is now further penalized in the region of the contact.

The qualitative discussion of the soliton's energetics given above is quite general and relies only on the existence of the soliton and the proximity effect. To illustrate our  arguments quantitatively we turn to a simplified,  analytically solvable model, based on the framework of the GL theory (\ref{GL general}). In it the effect of the patch is mimicked by a spatially dependent interband Josephson coupling $\kappa(x)$. The profile of this effective $\kappa(x)$ along the ring should be chosen to promote the alignment of the phases in the region of the patch. More precisely, we require 
\be
	 \kappa_{\rm in} > \kappa_{\rm out}
\label{kappa in out}
\ee
where $\kappa_{\rm in} \equiv \kappa(x)$ for $x$ in the patch and $\kappa_{\rm out} \equiv \kappa(x) $ otherwise. Notice that while for an $\spp$ ring $\kappa(x)$ has the same (positive) sign for all $x$, for an $\spm$ ring where $\kappa_{\rm out} <0$,  $\kappa(x)$ can, in the case of the strong proximity, reverse its sign in the patch area so that $\kappa_{\rm out} \kappa_{\rm in} <0$.

Before starting on the calculations we should discuss a potential caveat related to the use of the GL free energy expansion. It has recently been argued\cite{KS} that the accuracy of expansion (\ref{GL general}) which is controlled by the reduced temperature $(T_c - T)/T_c$ imposes a constraint $\psi_1 = \text{real number}\, \times \psi_2$ (see, however, Ref.~\cite{BabaevReply}). All other solutions, including the soliton one, although not forbidden in principle, should be obtained by supplementing expansion (\ref{GL general}) with higher-order terms\cite{Shanenko:2011} (cf.~Ref.~\cite{BabaevGLderivation}). We should point out that such argument is inapplicable in the presence of the proximity patch -- the main ingredient of our proposal. Indeed, it is quite clear that independent of a particular form of description the relative phase far from the patch is that of a bulk material while deep in the region of  the patch  $\alpha$ is controlled by the proximity effect as shown by a microscopic analysis\cite{AEK}. The existence of a (meta)stable soliton in our model relies only on an interpolation between these limits and is thus highly plausible.

 
To continue our analysis  we assume that the magnitudes of the order parameter $|\psi_i|$ is field- and coordinate-independent
which allows one to drop the potential energy $V_i(|\psi_i|)$. 
Let us introduce the following notation:
\be
	\lambda^{-2}_i  \equiv 8\pi e^2 |\psi_i|^2 /(m_i c^2), \quad \lambda^{-2}  \equiv \lambda_1^{-2} + \lambda_2^{-2}.
\ee
We focus on the limit of weak screening defined by $Rd/\lambda^2 \ll 1$. In this limit the difference between applied and total fluxes, as well as the difference between the Gibbs potential and the free energy, can be ignored.

\begin{figure}[t]
\centering
	\includegraphics[scale=0.54]{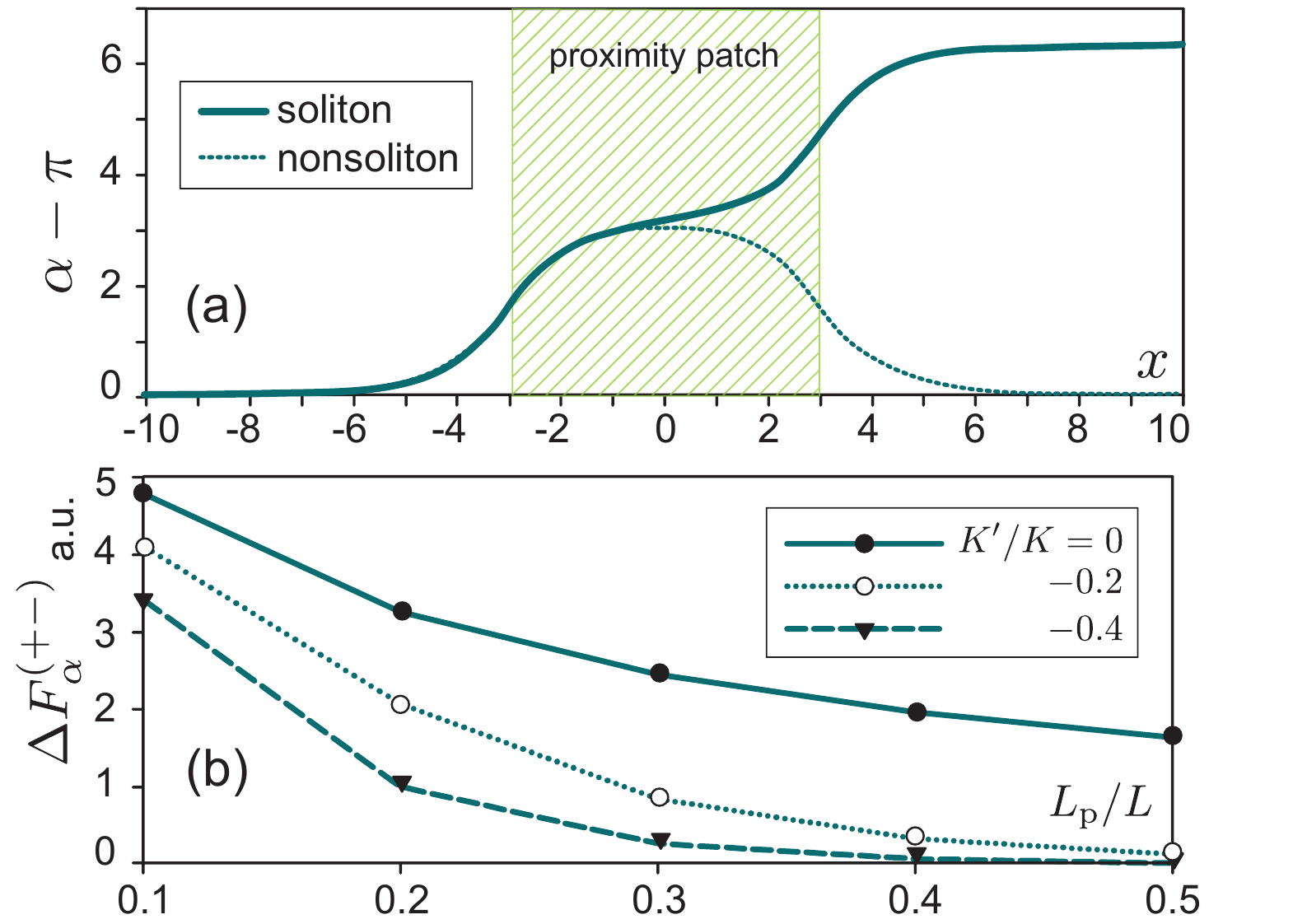}
	\caption{\label{soliton} (a) Spatial profile of the relative phase of the soliton and non-soliton solutions for $\spm$ pairing in the presence of the proximity patch. (b) Dependence of the excessive soliton's energy on the length of the proximity patch for several $K$. } 
	\vspace{-10pt}
\end{figure}

Following Ref.~\onlinecite{Kuplevakhsky:2011} we define new phase variables $\theta \equiv \lambda^2 (\theta_1\lambda_1^{-2}  + \theta_2\lambda_2^{-2} )$ and $\alpha \equiv \theta_1-\theta_2$. It is now possible to integrate out the $\theta$-dependence in Eq.~(\ref{GL general}) which leads to the following expression for the free energy of the ring\cite{fn3}:
\vspace{-4pt}
\begin{align}
	& F = F_\Phi + F_\alpha
	\label{F final}
	\\
	& F_\Phi  =  \eps_0 
	\big[ n_1(\lambda^2 /\lambda_1^2)+n_2  (\lambda^2/\lambda_2^2) - \Phi/\Phi_0 \big]^2
	\label{F Phi}
	\\
	& F_\alpha [\alpha]
	=
	 \eps_0   \, \frac{ R \lambda^4 }{ \pi \lambda_1^2 \lambda_2^2}
	 \int \! dx \, \Big\{ 
	\frac 1 2 (\pa_x \alpha)^2
	-
	K(x) \cos \alpha
	\Big\}
	\label{F alpha}
\end{align}
where $\Phi_0 \equiv hc/2e$, $n_i$ is an integer which describes the winding of $\theta_i$, $\eps_0  \equiv w d   \Phi_0^2/(16 \pi^2 R \lambda^2)$ is the electromagnetic energy scale and $K(x) \equiv \kappa(x) \, m_1 m_2 \Phi_0^2/(2 \pi^3 \hbar^4 \lambda^4)$ is the renormalized Josephson coupling. Minimizing $F_\alpha$ -- the flux-independent contribution  associated with the relative phase -- we find that $\alpha(x)$ is determined by the sine-Gordon equation with a spatially dependent mass:
\be
	\pa_x^2 \alpha(x) - K(x) \sin \alpha(x) =0
	\label{sine-Gordon}
\ee
subject to the boundary conditions specified by Eq.~(\ref{winding}).

At any given flux $\Phi$ thermodynamically stable state is the one which realizes the global minimum of $F$. It can be shown from Eqs. (\ref{F final})-(\ref{F alpha}) that in the absence of a proximity patch (i.e. when $K(x) = K_0 \equiv \text{const}$) the global minimum \emph {never} corresponds to a soliton state. 
While $F_\Phi$ itself allows for a stable soliton, its effect is countered by $F_\alpha$, which satisfies the following inequality:
\vspace{-3pt}
\be
	\Delta F_\alpha^{(0)} \geq \eps_0 / 4
	\label{constraint}
\ee
where $\Delta F_\alpha^{(0)} = (F_\alpha [\alpha]- F_\alpha[0])|_{K=K_0}$ is the difference between relative phase energies of a soliton and a conventional vortex solutions for $K(x) = K_0$. The lower bound for $\Delta F_\alpha^{(0)}$  is reached when  $\lambda_1 =\lambda_2$ and $K_0 =0$ i.e.~for a uniform phase winding. In this limit electromagnetic response of the soliton is equivalent to that of a half-quantum vortex \cite{Jang:2011}.

From now on we specialize on a physically reasonable setup in which lengths of both the patch and the soliton  are small compared to the perimeter of the ring\cite{fn4}. In this limit single-valuedness  condition (\ref{winding}) can be replaced with boundary conditions at infinity: $\alpha(\infty)=\alpha(-\infty)+2\pi$  for a soliton state and $\alpha(-\infty)=\alpha(\infty)$ otherwise. We have excluded states with larger winding numbers as they have higher energy. In terms of the winding numbers of each band the above conditions imply  $|n_1 - n_2| \leq 1$.

We now demonstrate that constraint (\ref{constraint}), which prevents the thermodynamic stability of the soliton, can be lifted in the presence of the proximity patch.  Detailed calculation of $K(x)$,  describing the full effect of the patch, requires the use of the microscopic theory and is beyond the scope of this paper. Generally speaking function $K(x)$ should be such that condition (\ref{kappa in out}) is satisfied. A straightforward analytical solution is available for a delta-patch model:
\be
	K(x) = K_0(1 - \gamma |K_0|^{-1/2} \delta(x-x_0))
	\label{delta-patch}
\ee
which describes a small, but very strongly coupled patch, parametrized by $\gamma$. For an $\spp$ ring we have $K_0>0$ and $\gamma <0$, while for a non-trivial $\spm$ pairing the opposite holds: $K_0<0$ and $\gamma >0$.

The presence of the $\delta$-function in Eqs.~(\ref{delta-patch}) and (\ref{sine-Gordon}) can be dealt with by imposing an additional boundary condition on the unperturbed ($K(x)=K_0$) solution obtained by integrating Eq.~(\ref{sine-Gordon}) around the location of the $\delta$-peak. We note that in this case a non-soliton solution, like a soliton one, may have a non-trivial structure in which  $\pa_x \alpha \neq 0$. Let us define $\Delta F_\alpha^{(i)}$ to be the difference in $F_\alpha$ evaluated between the lowest energy soliton and  non-soliton solutions for the ring whose pairing symmetry type is $i$. A straightforward calculation shows that
\begin{align}
	\phantom{aaaaaa} & \Delta F_\alpha^{(++)} = 
										8 F_{\alpha 0} , \quad \quad \quad \quad \quad \quad  \gamma \leq 0
	\label{deltaFspp}
\\
	\phantom{aaaaaa} & \Delta F_{\alpha }^{(+-)}= 2 F_{\alpha 0} \times
	 		 \left\{ \begin{array}{ll}
			4 - \gamma, & \quad 0\leq \gamma \leq 2
			 \\
			4/\gamma, & \quad \gamma\geq 2
	 \end{array} \right.
	 \label{deltaFspm}
\end{align}
where $F_{\alpha 0} \equiv  \eps_0 R  |K_0|^{1/2} \lambda^4/(\pi \lambda_1^2 \lambda_2^2)$. The above result is applicable when $R |K_0|^{1/2} \gtrsim 1$.

It follows from Eqs.~(\ref{deltaFspp}), (\ref{deltaFspm}) (in agreement with the earlier discussion) that the effect of the proximity patch on the winding of the relative phase is qualitatively different for $\spp$ and $\spm$ pairing symmetries. While in the former case the $\alpha$-winding of the soliton always costs non-zero energy $ \Delta F_\alpha^{(++)} $ limited from below by Eq.~(\ref{constraint}), for  $\spm$ symmetry the energy difference $\Delta F_{\alpha }^{(+-)}$ can be made arbitrary small, thus circumventing restriction (\ref{constraint}) and rendering the soliton thermodynamically stable. The independence of $\Delta F_{\alpha}^{++}$ on $\gamma$ is a consequence of the fact the soliton's kink and the patch ``repel" each other for $\spp$ pairing, and the lowest energy configuration corresponds to an infinite separation between them.

The conclusions derived from the delta-patch model are confirmed by numerical calculations for a finite length proximity patch. In this model the proximity is induced in a finite segment of length $L_{\rm p}$ such that $K(x) = K' = \text{const}$ on the patch and $K(x) = K \equiv \text{const}$ otherwise. We focus on the limit when the length of the soliton's kink is much smaller than the perimeter of the ring. The results of the calculations for $\spm$ pairing are presented on Fig.~\ref{soliton}. Fig.~\ref{soliton}a  shows the spatial profile of the relative phase for soliton and non-soliton solutions and Fig.~\ref{soliton}b demonstrates the reduction of the relative energy of the soliton as the length of the proximity patch is increased. 

%
\begin{figure}
\centering
	\includegraphics[scale=.8]{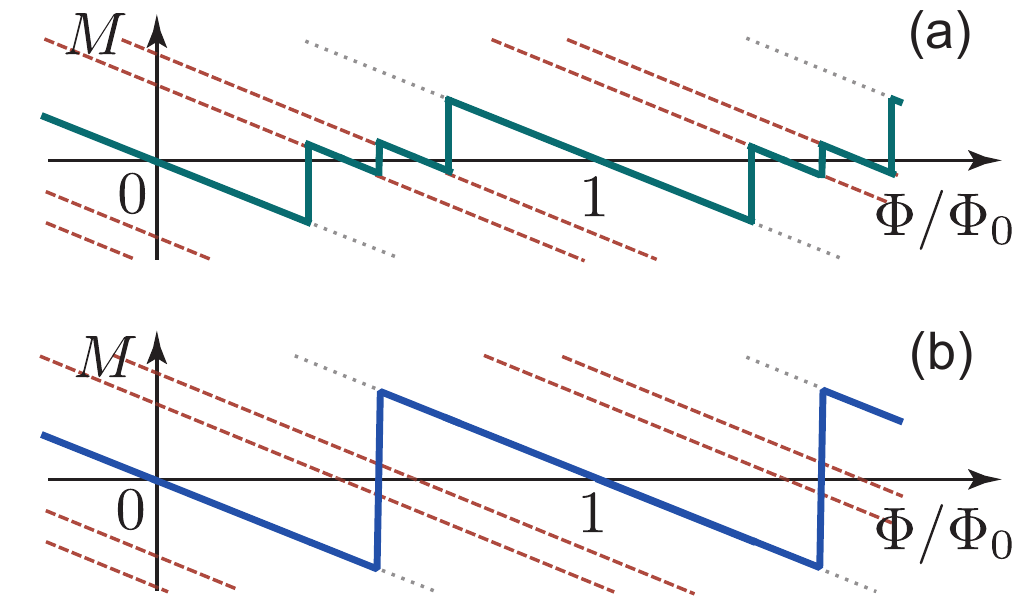}
	\caption{\label{magn} Magnetic moment  for stable, (a), and unstable, (b), soliton configurations in a ring geometry. Solid lines correspond to the equilibrium behavior. Dashed and dotted lines correspond to metastable soliton and vortex states respectively.} 
	\vspace{-15pt}
\end{figure}

\emph{Proposed experimental setup.} We now come to the question of experimental observation of  phase solitons. We consider the setup illustrated  on Fig.~\ref{patch}. To insure strong effect of the proximity the thickness of the ring in the direction perpendicular to the interface should be comparable or smaller than the coherence length.

There are several other factors  which would favor strong proximity effect \cite{BC1}. First, the $s$-wave SC has to have higher $T_c$. Second, the interboundary coupling between all the gaps has to be sufficiently strong. This implies not only  low boundary resistivity, but also requires the ratio of the normal state conductivities of the $s$ and $\spm$ materials to be much larger then one  - thus ensuring that the effect of the $s$-wave SC on the $\spm$ side is maximal \cite{BC2}.
Based on these requirement we suggest that a good candidate combination consists of a member of the 11 iron chalcogenide or 111 iron pnictide families (which are semimetals with $T_c$'s of about $10$K and typically have gaps with close values; some of these materials has been suggested to be $\spm$, Refs.~\onlinecite{Christianson:2008, Hanaguri:2010}), coupled to MgB$_2$ as an $s$-wave material. MgB$_2$ has  $T_c \approx 42$K and is a good metal in its normal state. Its own multigap nature should not be a problem in this context (provided it is in $s_{++}$ state).

In a ring geometry the phase soliton can be observed through measurements of ring's magnetic moment $M$. Fig.~\ref{magn} shows the dependence of $M$  on the applied flux $\Phi$ in the presence, (a), and in the absence, (b),  of a thermodynamically stable soliton, obtained from Eqs.~(\ref{F final}) - (\ref{F alpha}). A stable soliton state is seen  as two extra transitions between the adjacent vortex (fluxoid) states.  These two transitions correspond to a mismatched phase winding in the two bands $|n_1 - n_2|=1$ and are in general split because of the different band superfluid densities.

Even if the coupling to the proximity patch is not sufficient to induce the thermodynamic stability of the soliton, it can still be observed as a metastable configuration. Metastability of vortices is well-known\cite{Chen:2010} and can be a notorious problem. We suggest to look for metastable solitons through thermal cycling, which in practice can be realized by heating a part of the ring with a laser  pulse and then allowing the ring to cool down \cite{Tate:1989, Tate:1990}. To observe the soliton the system is repeatedly thermally cycled at a fixed value of the applied flux and a distribution of magnetic moments which correspond to different metastable states is collected. In the absence of solitons such distribution plotted at different values of flux should form a set of equidistant lines (dotted lines on Fig.~\ref{magn}, see also Fig.~1 in Ref.~\onlinecite{Tate:1990}). In the presence of a (metastable) soliton extra lines appear (dashed lines on Fig.~\ref{magn}); the relative number of points on these lines gives the probability of accessing a metastable soliton state\footnote{If the winding of the relative phase in the region of the proximity patch is associated with the suppression of the order parameter, then the slope of the soliton lines will be smaller than that of the vortex lines.}.

The metastability of a non-equilibrium state can be greatly improved for a thin ring, where dynamics of the order parameter  is one-dimensional. In realistic experimental settings the probability of accessing a metastable state with an excess energy $E$ separated by the barrier $V$ from the ground state is of the order of  $e^{- x E /V} $ where $x$ is a parameter which depends logarithmically on the product of cooling time and characteristic attempt
 frequency. In the case of a thin ring interrupted by a weak link dominant contribution to $E$ is of the electromagnetic origin so that $ E \sim \Phi_0/R^2$. At the same time the dominant contribution to $V$ is set by the weak link's energy making $V$ only weakly $R$-dependent. Hence to observe metastable states the use of larger rings is desirable.

Although the detection of the soliton for an $\spp$ ring in the presence of the $s$-wave proximity patch is highly unlikely, it is nevertheless a possibility. To rule out such scenario the dependence of the stability region $\Delta \Phi_{\rm sol}$ (for the thermodynamically stable soliton)  or of the probability of accessing a soliton through thermal cycling $p_{\rm sol}$ (for a metastable soliton) should be checked for different lengths of the patch. In the $\spm$ case the soliton's energy is reduced as the length  is increased (Fig.~\ref{soliton}b) which leads to the growth of $\Delta \Phi_{\rm sol}$ and $p_{\rm sol}$, while for $\spp$ it, at best, remains constant. The observation of such growth would unambiguously imply the $\spm$ pairing.

To conclude, we have considered the influence of the proximity effect on the stability of a phase soliton in a two-band superconductor. We have shown that the proximity between $\spm$ and $s$-wave superconductors can reduce the energy of the soliton and suggested a practical experimental setup which utilizes this observation to distinguish between $\spm$ and $\spp$ types of pairing.

\textit{Acknowledgments} VV and VS would like to thank Jasper van Wezel, Thomas Prolier, Alexei Koshelev and Michael Norman for useful discussions. The financial support was provided  by the Center for Emergent Superconductivity, an Energy Frontier Research Center funded by the U.S.~DOE, Office of Science, under Award No.~DE-AC0298CH1088.


\begin{thebibliography}{1}
\bibitem{fn1}We use the convention in which $2m$ is the mass of the Cooper pairs and $|\psi|^2$ is their density.
\bibitem{fn2} Technically speaking in a ring geometry we are dealing with coreless fluxoid states. For the sake of convenience we will refer to a state with non-zero phase winding around the ring as a vortex state ($n_\alpha =0$) and as a soliton state ($n_\alpha \neq 0$).
\bibitem{fn3} This expression, strictly speaking, is applicably only to an infinitely long cylinder. It can be shown however that it also applies for a ring  in which case the combination $Rd/\lambda^2$ should be replaced with an expression which depends on the inductance of the ring.
\bibitem{fn4} These assumptions are not crucial and can be relaxed at the expense of mathematical clarity.
\end{thebibliography}


\end{document}